\documentclass[runningheads]{llncs}
\usepackage[T1]{fontenc}
\usepackage{graphicx}
\usepackage{todonotes}

\usepackage{url} 

\usepackage{subcaption}
\usepackage{booktabs}
\usepackage{tabularx}

\usepackage{paralist}

\begin{document}
\title{Opportunities and Limitations of GenAI in RE: Viewpoints from Practice}
%
%
\author{Anne Hess\inst{1}\orcidID{0000-0001-7614-5802} \and
Andreas Vogelsang\inst{2}\orcidID{0000-0003-1041-0815} \and
Xavier Franch\inst{3}\orcidID{0000-0001-9733-8830} \and
Andrea Herrmann\inst{4}\orcidID{0000-0002-4234-8422} \and
Sylwia Kopczyńska\inst{5}\orcidID{0000-0002-9550-3334} \and \\
Alexander Rachmann\inst{6}\orcidID{0000--0001-8217-7149} 
}
\authorrunning{A. Hess et al.}
%
\institute{University of Applied Sciences Würzburg-Schweinfurt, Würzburg, Germany 
\email{Anne.Hess@thws.de}\\
\and
University of Duisburg-Essen, Essen, Germany\\
\email{andreas.vogelsang@uni-due.de}\\
\and
Universitat Politècnica de Catalunya, Barcelona, Spain\\
\email{xavier.franch@upc.edu}\\
\and 
Herrmann \& Ehrlich, Stuttgart Germany\\
\email{herrmann-ehrlich@gmx.de}\\
\and
Poznan University of Technology, Poznań, Poland\\
\email{sylwia.kopczynska@cs.put.poznan.pl}\\
\and
Hochschule Niederrhein, Krefeld, Germany\\
\email{Alexander.Rachmann@hs-niederrhein.de}
}
\maketitle              
\begin{abstract}
\textbf{Context and motivation:}
With the rapid advancement of AI technologies, there is an increasing need to understand how AI can be effectively integrated into RE processes. In recent years, several studies have explored the potential and challenges of applying GenAI to support or even automate RE-related activities. 

\textbf{Question/problem:} 
Despite the existing body of knowledge on AI's potential for supporting RE activities, there is limited evidence on its practical applicability and limitations from an industry perspective. 

\textbf{Principal ideas/results:} 
To address this gap, we conducted a survey with RE practitioners in collaboration with the IREB Special Interest Group on AI \& RE. In addition to describing our research methodology and survey design, we present insights from our quantitative and qualitative data analyzes. These insights include practitioners' perspectives on current usage scenarios, concerns, experiences—both positive and negative—as well as training needs related to using GenAI in requirements elicitation, analysis, specification, validation, and management.

\textbf{Contribution:} 
This study provides empirical evidence on the practical use of GenAI in RE, offering insights into its benefits, challenges, and training needs. The findings inform future research and industry strategies, guiding effective AI integration and skill development for improved RE processes and results.

\end{abstract}
%
%
\section{Introduction}

Requirements Engineering (RE) is a crucial discipline in software development, which provides the foundation for successful project results by ensuring that stakeholder needs are precisely elicited, analyzed, specified, validated, and managed~\cite{ref_ASJ2020,pohl2016requirements}.
In recent years, the integration of Artificial Intelligence (AI) into RE processes has emerged as a promising approach to improve both the efficiency and effectiveness of performing the aforementioned RE-related activities~\cite{ref_FN2020,ref_LRR2022,HIT2024,11190349,10629163}.
Research in this area spans a wide range of AI-based approaches, supporting requirements elicitation and classification \cite{ref_AP2025,White2024}, requirements specification \cite{11122187}, or requirements prioritization \cite{10360116,11082164}. 

Beyond these contributions, a growing number of empirical studies, particularly literature reviews, have examined how AI is being applied within RE, reflecting the increasing interest in leveraging AI techniques to support and even automate various RE-related activities \cite{ref_CC25,ref_CHL25,ref_KDS24,ref_VR25,ref_NCP25,ref_ZDA25}.
These studies highlight both the potential benefits and challenges of integrating AI into RE processes, contributing to a growing and diverse body of work that evolves with advancements in AI technologies. Notably, the research suggests that although AI can accelerate RE activities and improve result quality, it also raises concerns regarding the reliability and semantic accuracy of AI-generated artifacts, reinforcing the necessity of continued human oversight. 

While knowledge on AI’s potential and limitations in RE exists, there is limited evidence concerning its practical applicability from an industry perspective. To address this gap, we conducted an online survey among RE practitioners to investigate their use and perceptions of Generative AI (GenAI) in professional contexts. Our analysis of the elicited quantitative and qualitative data offers insights into typical usage scenarios, concerns, and threats impeding GenAI adoption in RE. It also highlights both positive and negative experiences associated with GenAI in RE-related activities and identifies training needs.

This study was carried out as part of our  collaboration within the Special Interest Group on Artificial Intelligence \& Requirements Engineering of the International Requirements Engineering Board (SIG \#AIREB\footnote{\url{https://ireb.org/en/community/special-interest-group/sig-aireb}, last access October 23rd, 2025}). The group is dedicated to exploring how AI technologies can enhance and transform RE, aiming to establish standards, identify best practices, and offer guidance and training to responsibly integrate AI into RE processes. 

The remainder of this paper is structured as follows: Section~\ref{sec:Methodology} describes the methodology and design of the study. Section~\ref{sec:Results} presents the results obtained from both quantitative and qualitative data analysis. In Section~\ref{sec:Discussion} we discuss our findings in relation to existing research  and potential threats to validity. Finally, Section~\ref{sec:Conclutions} concludes the paper with a summary of the key insights and outlines directions for future work.

\section{Study Design and Methodology}
\label{sec:Methodology}


\noindent\textbf{Research Objective and Research Questions.}
As motivated in the introduction section, our study aimed to explore how RE practitioners use and perceive GenAI in their professional activities. To achieve this goal, we derived the following research questions.

\begin{compactitem}
    \item RQ1: How frequently do RE practitioners use GenAI in their work and for which RE activities?
    \item RQ2: What factors prevent RE practitioners from using GenAI, and what concerns do they have about its use in RE? 
    \item RQ3: What positive and negative experiences have RE practitioners encountered when using GenAI in RE and how do they assess its usefulness and harmfulness?
    \item RQ4: In which areas do RE practitioners see a need for additional training on GenAI?  
\end{compactitem}

To address these research questions, we conducted an online questionnaire-based survey among practitioners. This cost-effective method allowed us to collect and analyze data from a large sample, providing an overview of GenAI usage in RE. The study was designed according to the guidelines proposed by Wohlin~et~al.~\cite{Wohlin_etal_2012} and Molléri~et~al.~\cite{molleri2020empirically}.


\noindent\textbf{Survey instrument.} Our questionnaire comprised a total of 37 questions (both open and closed questions), organized into several sections:
\begin{compactitem}
    \item \textbf{Introduction} outlined the survey’s objectives, relevant policies, estimated completion time, and provided contact information of the research team; 
    \item \textbf{Demographics and Experience} gathered data on participants’ professional experience both overall and in RE, their organization's size, current role, industry sector, country, experience with specific RE activities, and previous exposure to GenAI tools like ChatGPT;
    \item \textbf{Use of GenAI in RE} investigated its application across various RE activities, including elicitation, analysis, specification and modeling, validation and quality assurance, and management. In addition to gathering insights on positive and negative experiences, we identified reasons and barriers that prevented participants from using GenAI in RE. We also collected data on the perceived usefulness, as well as the limitations and risks associated with GenAI in each RE activity and related tasks/usage scenarios;
    \item \textbf{Training Needs} questioned respondents whether they would like to receive additional training, preferred training formats, and the skills perceived as most essential for effectively adopting AI in RE; 
    \item \textbf{Closing section} asked for final remarks and optional comments.
\end{compactitem}


The survey questions were developed and validated through an iterative process. The core objectives, along with the initial topics, ideas, and concepts, were introduced and discussed during one of our regular meetings within the SIG \#AIREB. Building on these discussions, the first two authors drafted an initial structure and a set of questions, which were presented at a subsequent follow-up meeting. Additionally, the initial questionnaire was distributed by email to all 35 members of the SIG, as not all members could attend the meetings. Feedback gathered during the meeting and via email from members (including input from the other four authors) was incorporated into the set of questions, which was then implemented on the LimeSurvey platform\footnote{\url{https://www.limesurvey.org/}, last access on January 18, 2026}, chosen for its accessibility, usability, and security features. The survey was set up to collect anonymous responses. Before its public distribution, the survey underwent a final validation round and was pilot-tested again within the SIG \#AIREB to ensure quality and clarity. This final validation led to minor adjustments in wording, spelling, and question order. No major changes were necessary following the pilot study, and the finalized questionnaire (see link above) was used in the data collection phase. All responses collected during the final validation were deleted and not included in the analysis phase.

\noindent\textbf{Target Population and Sample.}
We defined the target population as participants in software development projects and product teams who have experience in RE. We aimed to obtain a diverse sample including practitioners from various countries, domains, and organizational contexts, thereby reflecting the diversity of RE practice in industry. 

\noindent\textbf{Distribution Strategy.} Since no single distribution channel provides direct access to a representative sample of the target population, we employ multiple online dissemination strategies. The survey was distributed through: 1) direct contacts of the authors and an author's newsletter, 2) LinkedIn social media accounts of both the SIG \#AIREB and authors, 3) IREB newsletter, and 4) the mailing list of the RE Specialist Group of the German Informatics Society (mid-February 2025).

Data collection occurred between November 1, 2024, and March 31, 2025. Due to public invitations, the exact response rate can't be calculated, but with about 130 participants starting the survey and 57 completing it, the response rate is considered low (which is typical for online surveys).

\noindent\textbf{Analysis Strategy.} In this analysis activity, all authors were involved. We conducted both quantitative and qualitative analysis. For closed-ended questions, such as multiple-choice and Likert-scale items, we applied descriptive statistics and frequency analysis. For open-ended questions, we performed a manual thematic analysis, systematically reviewing textual responses to identify recurring codes, themes, and patterns related to our research questions, particularly positive and negative experiences (RQ3). Two authors developed the initial coding schema after several consolidation meetings and pilots. This schema was reviewed by two other authors, which led to a revised iteration. The final coding schema, detailed in our codebook, includes 14 themes and 52 codes (see the shared material linked above for more information). 

Before starting data analysis, we screened all responses to the survey. Although only 57 participants completed the survey, we decided to include all data in the analysis, as we believe the presence of unanswered questions does not compromise the validity of the data. Unfortunately, we were unable to reliably interpret the reasons for these omissions, as they could stem from various factors, such as participants' lack of confidence due to limited experience with GenAI or fatigue from the lengthy survey questions. We also retained responses from participants with lower RE expertise, as they represent potential target groups for future RE training. For traceability, during our thematic analysis, we explicitly marked the responses of the participants who did not complete the survey.


\section{Results}
\label{sec:Results}
\subsection{Participants and Demographics}
According to our records in the Limesurvey tool, 130 people started the survey via the link that we shared via various channels and shared demographics and other data, but only 57 completed the survey (see also previous Section~2). 
%
The majority of our participants are employed in industry, particularly in large enterprises (74), medium enterprises (16), small enterprises (8), and micro-entreprises (6). The other respondents characterized their organization as research-technology transfer (8), and as university/research (6). 
There was a diversity of application domains in which the participants worked over the past 5 years, but more than half of the participants worked in the IT / Software domain (84). Moreover, 
 most of the participants indicated a professional experience of more than 10 years in the RE discipline (63), followed by 6-10 years (26), and 3-5 years (21). A total of 17 people indicated an experience level of less than 2 years or being new to the topic of RE. Although we aimed for geographic diversity, most of the respondents were from European countries (116). Few participants were from Asia (9), North America (4), South America (3), Africa (2), and Australia or New~Zealand (1).

\subsection{RQ1: Utilization of GenAI in RE}

\begin{figure}
    \centering
    \includegraphics[width=1\textwidth]{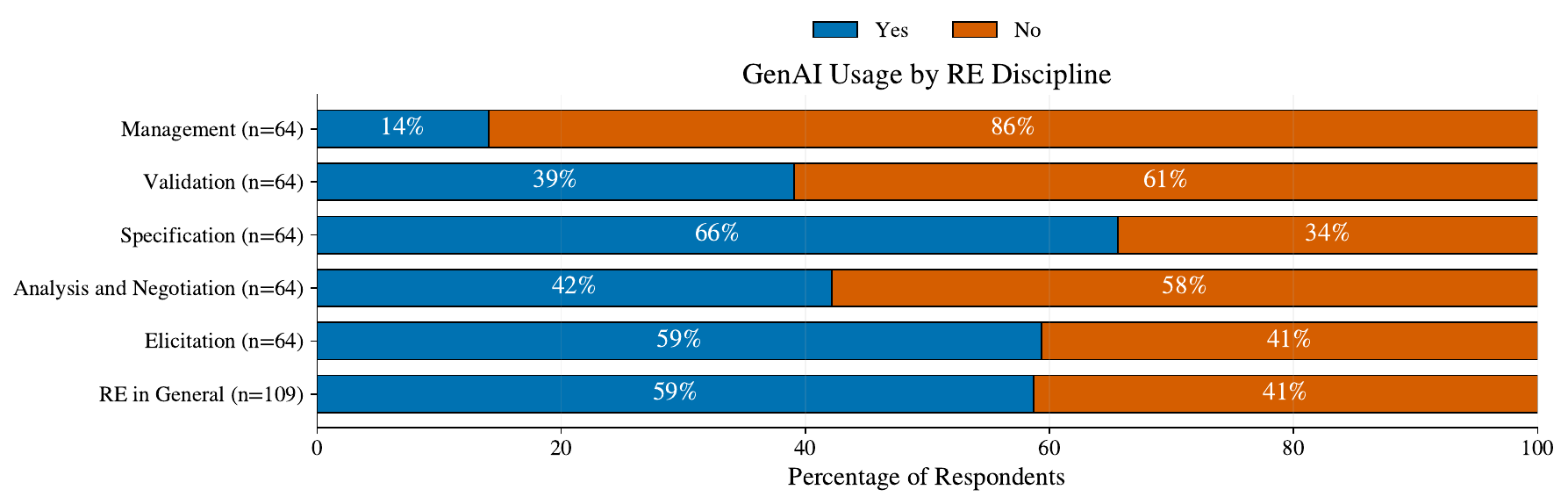}
    \caption{Have you used GenAI for RE?}
    \label{fig:usage}
\end{figure}

Figure~\ref{fig:usage} shows the percentage of respondents who have already applied GenAI in RE in general and for specific RE activities. The figure illustrates that over half of the respondents have utilized GenAI in RE, mainly for specification and elicitation. GenAI was applied the least frequently in requirements management.

\subsection{RQ2: Barriers, Limitations and Threats}
\label{sec:Threats}

\begin{figure}
    \centering
    \begin{subfigure}[t]{0.9\textwidth}
        \centering
        \includegraphics[width=\textwidth]{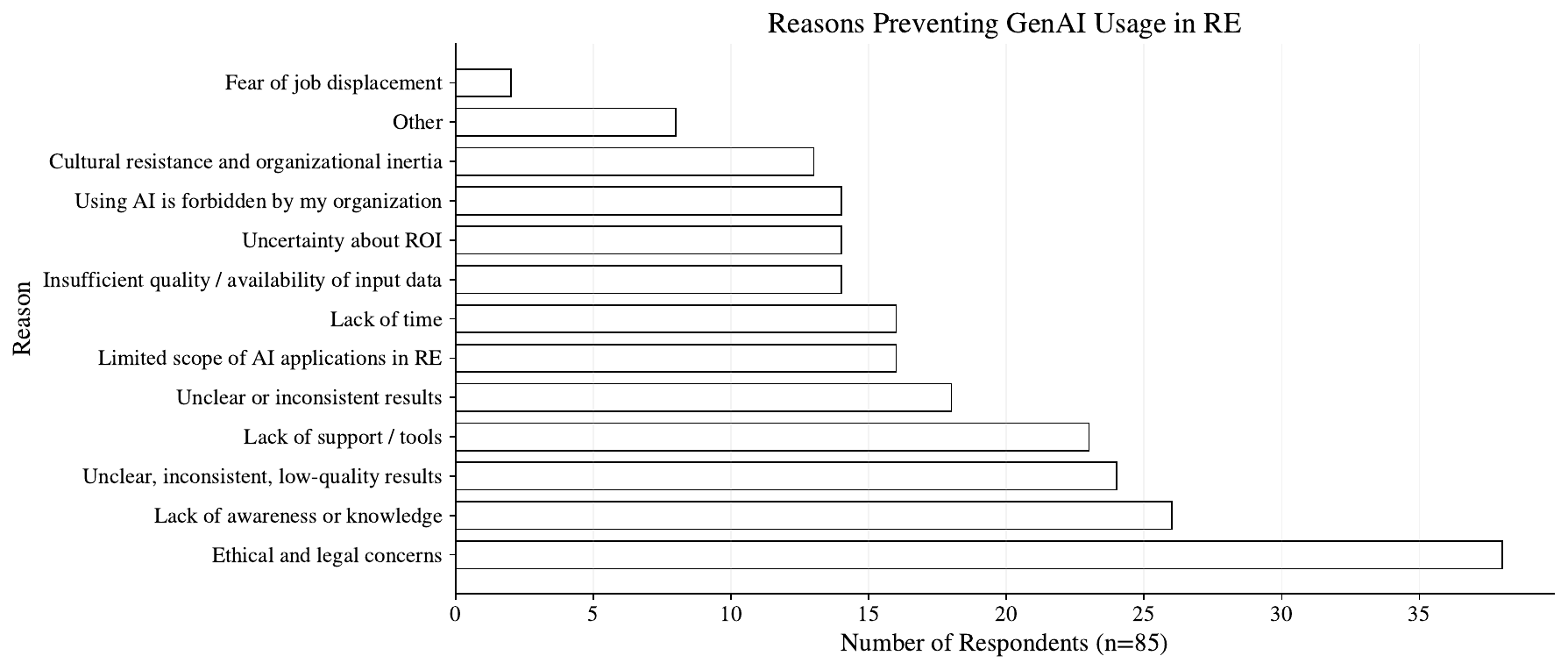}
        \caption{What reasons prevent(ed) you from using GenAI in RE?}
        \label{fig:prevented}
    \end{subfigure}
    \hfill
    \begin{subfigure}[t]{\textwidth}
        \centering
        \includegraphics[width=\textwidth]{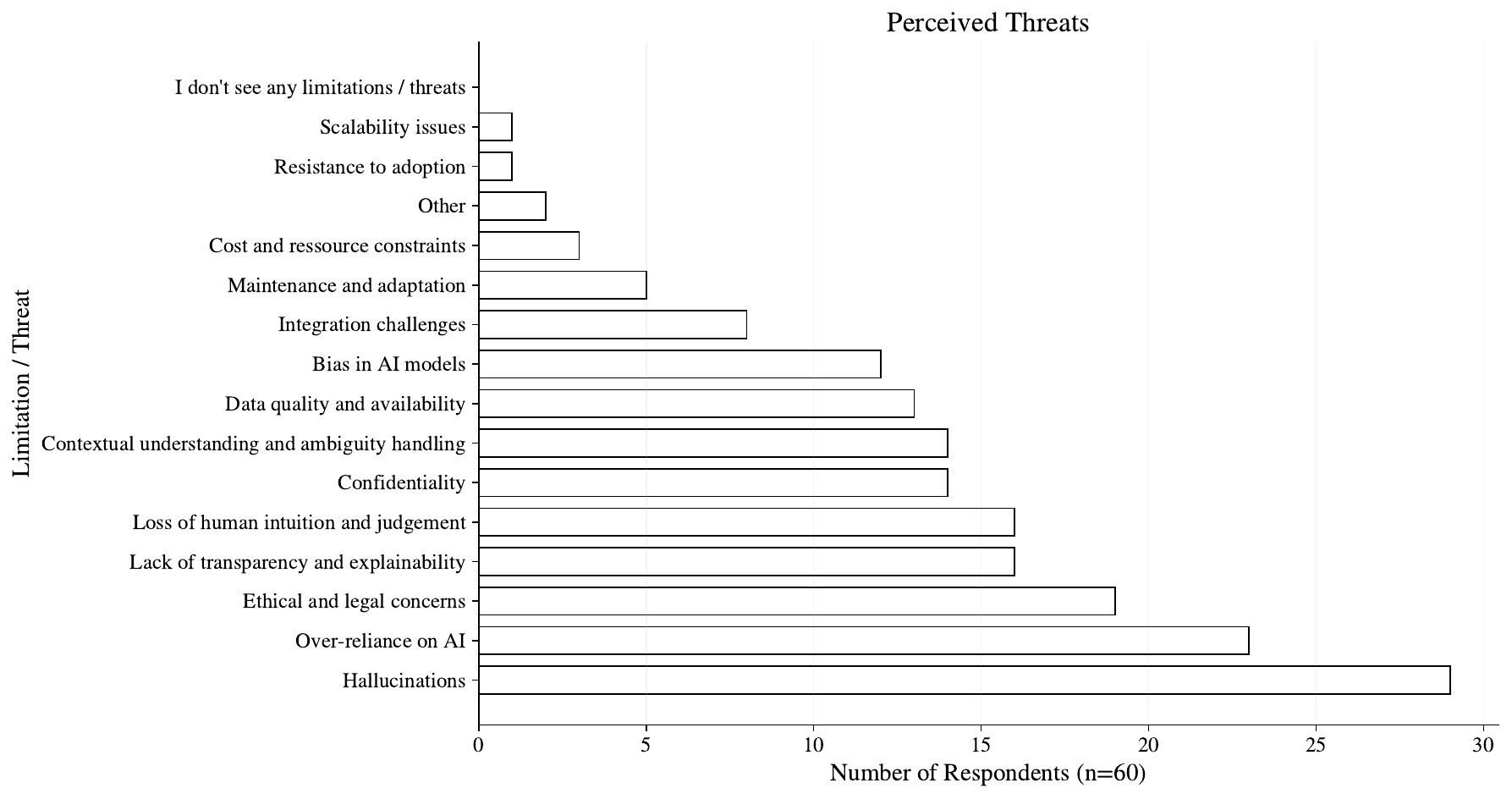}
        \caption{Select up to three limitations/threats that you see in GenAI in RE.}
        \label{fig:threats}
    \end{subfigure}
    \caption{Bar charts showing (a) reasons that prevented respondents from using GenAI in RE and (b) perceived threats and limitations.}
    \label{fig:prevention_threats_combined}
\end{figure}

For participants who had not yet used GenAI in RE or  encountered circumstances preventing its use, we further explored the underlying reasons. As Figure~\ref{fig:prevented} shows, ethical and legal concerns were the most frequently reported barriers, followed by lack of awareness or knowledge, insufficient organizational support, and low-quality results. Additionally, Figure~\ref{fig:threats} summarizes the threats and limitations perceived by all respondents when using GenAI in RE. Among these, hallucinations and over-reliance on AI were identified as the most frequent issues.

\subsection{RQ3: Applying GenAI in RE - Experience, Usefulness and Harmfulness}


\textbf{Requirements Elicitation.}
As shown in Figure~\ref{fig:usage}, elicitation is one of the two activities where more than half of the respondents have already used GenAI. 
We asked these respondents about their positive and negative experiences using GenAI in the context of elicitation. Thematic analysis of their comments revealed that GenAI is most beneficial for ``generation'' tasks or ``NLP'' tasks like generating requirements (e.g., based on interview results), formulating questions (e.g., for interview preparation), and summarizing content in large documents. Some respondents compared GenAI to ``roles'' like a sparring partner or assistant, particularly valuable during brainstorming sessions or when serving as a domain expert. Respondents generally found GenAI to be the most advantageous when supporting easy activities, with time savings being the most cited benefit, followed by output quality and output completeness.
However, output quality was also highlighted as a significant drawback, particularly due to hallucinations and overly simplistic results. These issues lead to human-experienced challenges, such as the need for human reviews, lack of trust, and potential time waste. 

\begin{figure}
  \centering

  \begin{subfigure}[t]{1\textwidth}
    \centering
    \hspace*{4cm}%
    \includegraphics[width=0.5\textwidth]{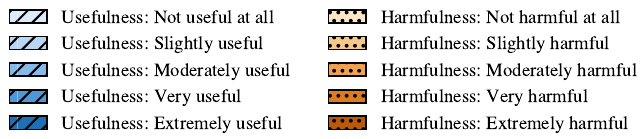}
  \end{subfigure}

  \vspace{1em} 

  \begin{subfigure}[t]{1\textwidth}
    \raggedleft 
    \includegraphics[width=1\textwidth]{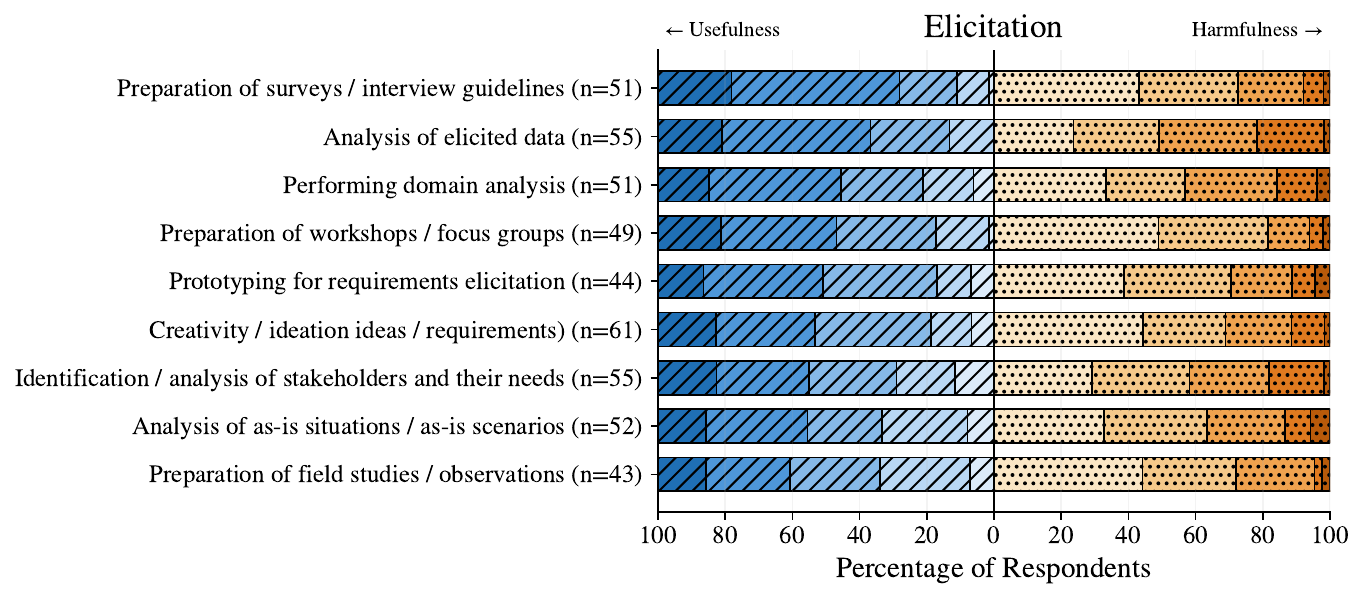}
    \caption{Elicitation}
    \label{fig:elicitation-assessment}
  \end{subfigure}

  \vspace{1em} 

  \begin{subfigure}[t]{1\textwidth}
    \raggedleft
    \includegraphics[width=1\textwidth]{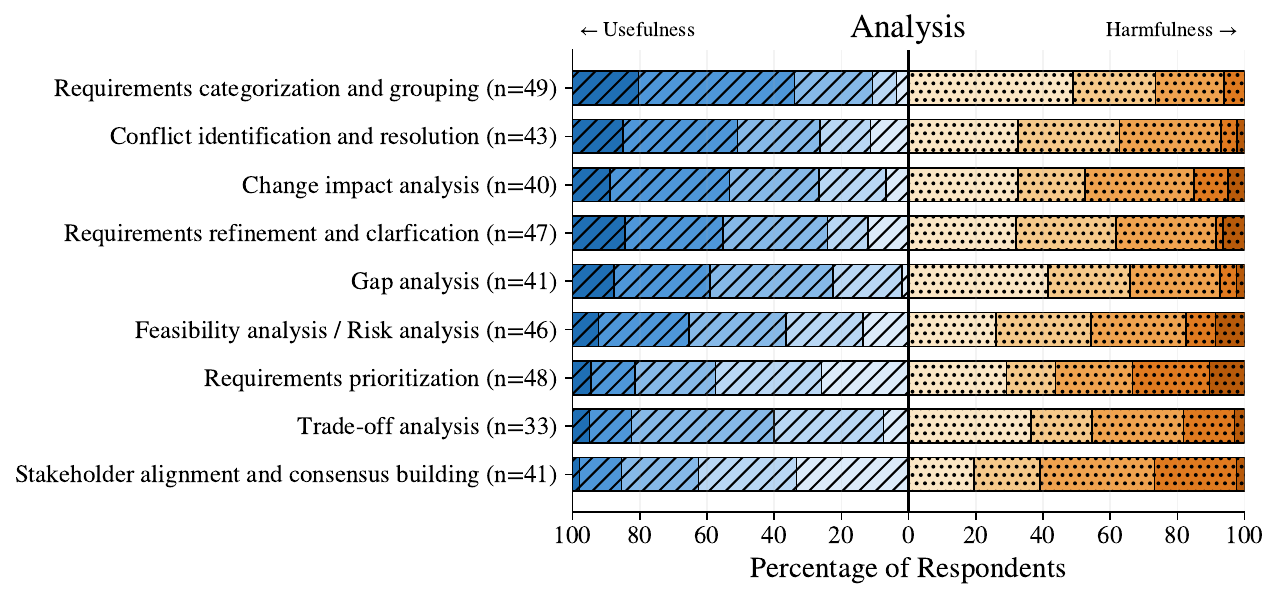}
    \caption{Analysis and Negotiation}
    \label{fig:analysis-assessment}
  \end{subfigure}

  \vspace{1em} 

  \begin{subfigure}[t]{1\textwidth}
    \raggedleft
    \includegraphics[width=1\textwidth]{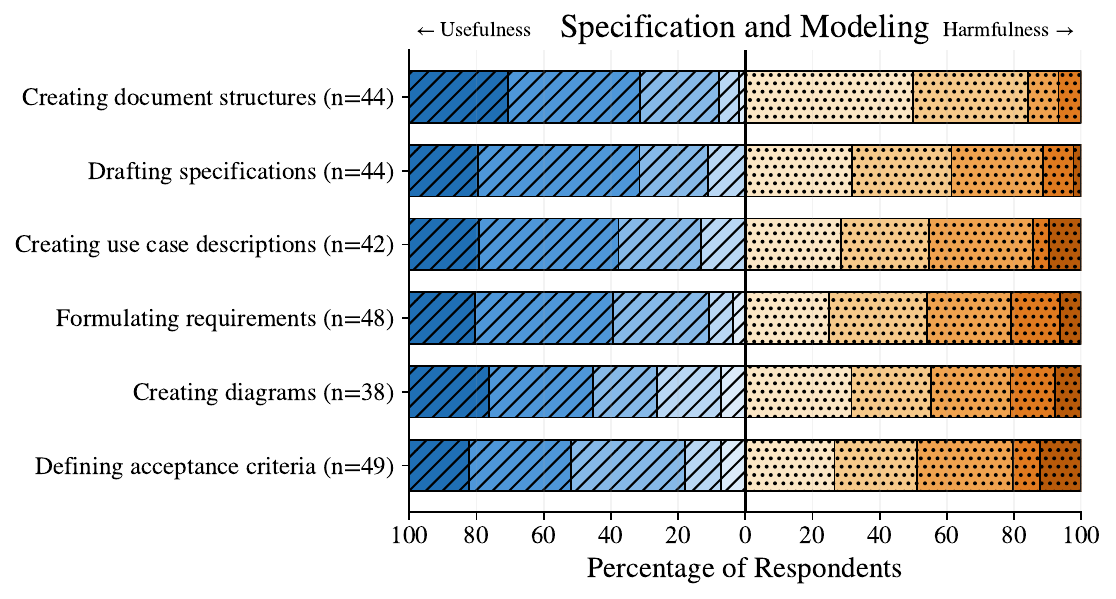}
    \caption{Specification and Modeling}
    \label{fig:specification-assessment}
  \end{subfigure}

  \caption{How useful/harmful is GenAI for certain RE tasks?}
  \label{fig:useharm-all}
\end{figure}

These findings are consistent with the observations derived from the quantitative data analysis. Figure~\ref{fig:elicitation-assessment} shows how practitioners assess the usefulness and harmfulness of GenAI for specific tasks within requirements elicitation. 
The results show that most of the respondents had positive experiences with GenAI, particularly in preparation tasks such as preparing surveys, workshops, or field studies. For these tasks, respondents assessed GenAI to be rather useful and cause little to no harm. Using GenAI for data analysis was evaluated as useful, but it also seems to have some risks.  

\noindent\textbf{Requirements Analysis and Negotiation.}
As shown in Figure~\ref{fig:usage}, analysis and negotiation exhibit relatively low GenAI, with around 58\% of the respondents not having used it. Those who did apply GenAI in this area found it useful for ``requirements analysis'', in particular for identifying conflicts between requirements and enhancing or reformulating them. However, respondents emphasized that such benefits require users' knowledge of GenAI and effective prompting. As with elicitation, time consumption was noted as a major drawback, largely due to the need for human reviews stemming from issues such as hallucinations and unusable outputs.

Figure~\ref{fig:analysis-assessment} presents how practitioners assess the usefulness and harmfulness of GenAI for specific analysis and negotiation tasks. The results show that respondents generally find GenAI helpful for tasks such as requirements categorization and grouping, gap analysis, where usefulness ratings are high and harmfulness is perceived as low. However, tasks that involve trade-off analysis or stakeholder alignment and consensus building are seen as more challenging: while some usefulness is recognized, respondents also report a higher potential for harm, indicating that these tasks may require more caution and human oversight.

\noindent\textbf{Requirements Specification and Modeling.}
Specification is the activity where GenAI has been adopted by 66\% of the respondents (Figure~\ref{fig:usage}), which is the largest share among all activities. The thematic analysis of comments on positive and negative experiences highlighted GenAI's benefits in ``model management'' tasks, such as model creation and translation, and ``modeling process'' tasks like suggesting missing aspects. Its support for ``generation'' tasks, particularly in generating acceptance criteria, was also noted. Respondents frequently mentioned time savings and improved output quality as key advantages. However, drawbacks included output quality issues such as hallucinations, incorrect solutions, dubious acceptance criteria, and inappropriate modeling suggestions, all of which require thorough proofreading.

As shown in Figure~\ref{fig:specification-assessment}, respondents perceive GenAI to be particularly useful for creating document structures and drafting specifications, with low levels of perceived harmfulness. Tasks such as formulating requirements, creating diagrams, and defining acceptance criteria are also overall positively evaluated, though with slightly more mixed experiences. The results suggest that practitioners see value in GenAI as a support tool for producing and structuring specification content, but that its application still requires careful review to avoid quality issues.

\noindent\textbf{Requirements Validation and Quality Assurance.}
Figure~\ref{fig:usage} shows that validation is an RE activity with relatively little reported usage of GenAI: 39\% of respondents have already applied it in this context. Compared to the previously discussed RE activities, we received fewer comments on positive and negative experiences. The analysis revealed that GenAI was especially beneficial for ``validation planning'' tasks, such as identifying quality criteria, and for ``property/concept validation'' tasks, like ensuring requirements' consistency and adherence to standards or norms. Although drawbacks such as hallucinations and the need for human reviews were noted, they were mentioned less frequently than in other activities.   

\begin{figure}
  \centering

  \begin{subfigure}[t]{\textwidth}
    \centering
    \hspace*{4cm}%
    \includegraphics[width=0.6\textwidth]{figures/legend.pdf}
  \end{subfigure}

  \vspace{1em} 

  \begin{subfigure}[t]{1\textwidth}
    \raggedleft
    \includegraphics[width=1\textwidth]{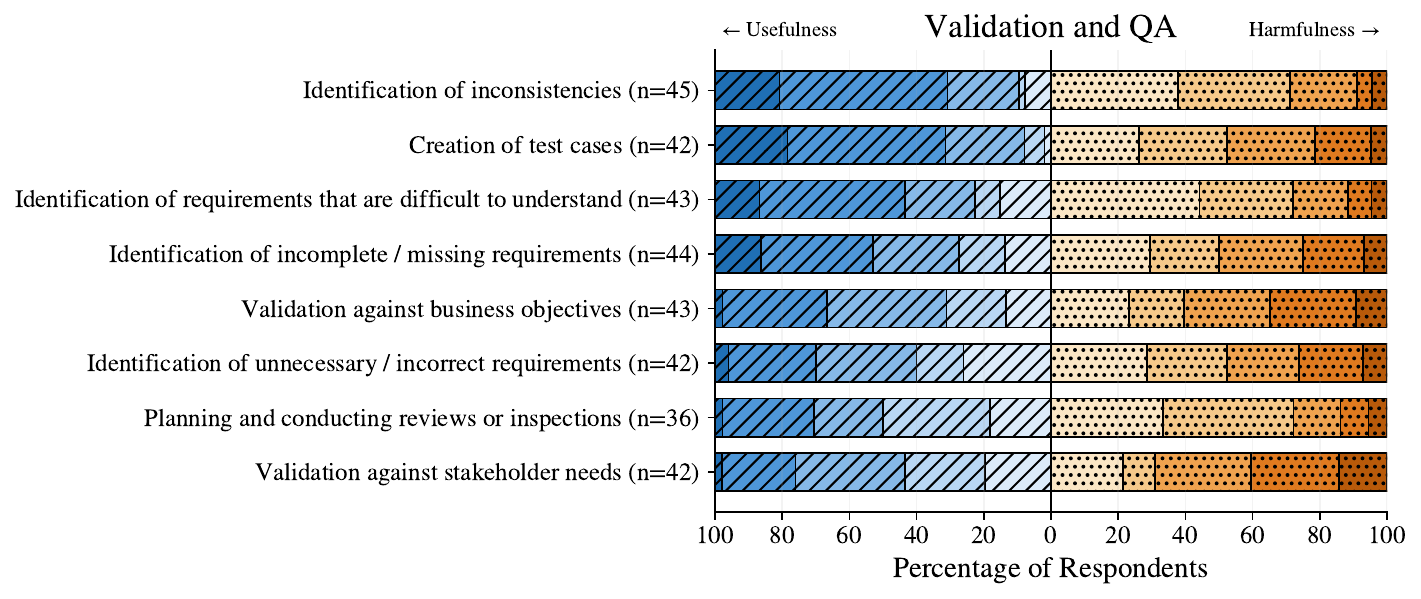}
    \caption{Validation and QA}
    \label{fig:validation-assessment}
  \end{subfigure}

  \vspace{1em} 

  \begin{subfigure}[t]{1\textwidth}
    \raggedleft
    \includegraphics[width=1\textwidth]{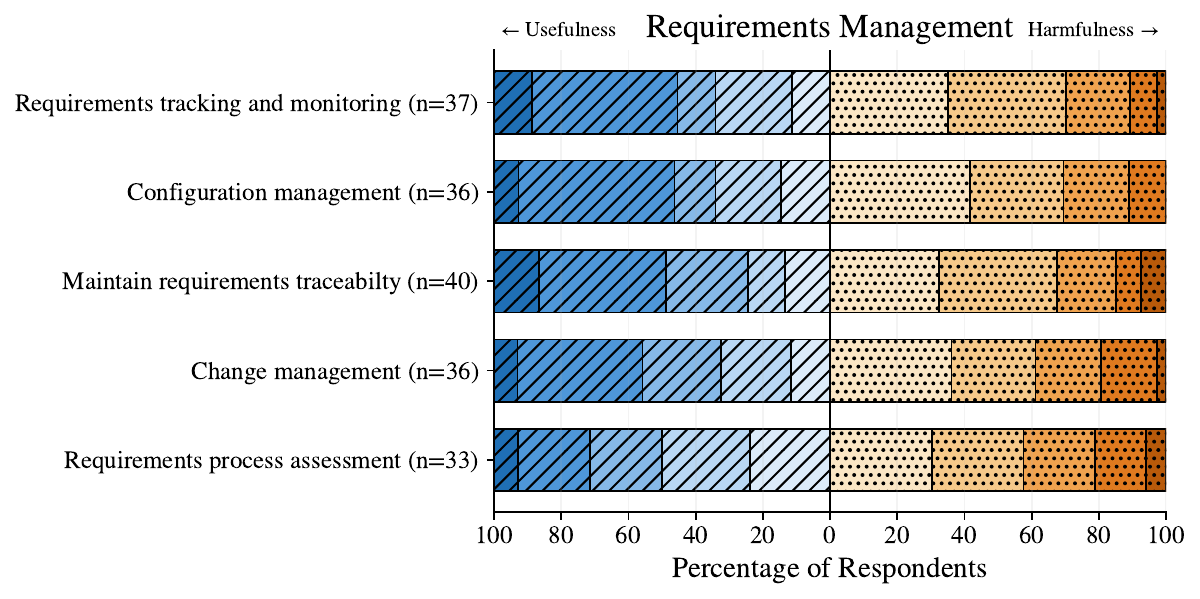}
    \caption{Requirements Management}
    \label{fig:management-assessment}
  \end{subfigure}

  \caption{How useful/harmful is GenAI for certain RE tasks?}
  \label{fig:useharm-all-2}
\end{figure}

The analysis of the quantitative data illustrated in Figure~\ref{fig:validation-assessment} shows that GenAI is considered especially useful for tasks such as identifying inconsistencies, detecting incomplete or unclear requirements, and creating test cases, which receive high usefulness ratings and low perceived harmfulness. More complex validation activities, such as validation against stakeholder needs or planning and conducting reviews, receive more mixed evaluations. While some practitioners find GenAI helpful, others see potential risks, suggesting that these activities still rely heavily on human expertise and judgment.

\noindent\textbf{Requirements Management.}
Management is the activity with the lowest reported GenAI usage, with only 14\% of the respondents indicating previous use (Figure~\ref{fig:usage}). Thematic analysis of their comments did not provide significant information due to the limited number of responses. 
The quantitative data analysis, illustrated in Figure~\ref{fig:management-assessment}, indicates that while GenAI is seen to be moderately useful for tasks such as requirement tracking, maintaining traceability and configuration management, its perceived usefulness is less pronounced compared to other activities, although the harmfulness ratings are low. More strategic or process-related tasks, such as change management or assessment of requirements processes, are considered less suitable for GenAI support. In general, the results suggest that the role of GenAI in the management of requirements is currently limited, likely due to the complexity and organizational nature of these tasks.

%

\subsection{RQ4: Training and Skills}

A large part of our respondents (90\%) agreed that the set of skills of requirements engineers will need to change as AI becomes more prevalent in RE. 

The survey responses highlight that requirements engineers will increasingly need skills in AI literacy, prompt engineering, and critical evaluation. Many emphasized the importance of understanding how AI works, its limitations, and how to effectively use AI tools to support, but not replace, their work. Prompting skills were repeatedly mentioned, along with the ability to validate and critically assess AI-generated outputs, given risks of errors, biases, or hallucinations. Respondents also noted a shift in focus from detailed documentation to higher-level skills such as stakeholder management, communication, risk awareness, and ethical considerations. In general, requirement engineers must adapt by learning to collaborate with AI responsibly, ensuring that human expertise remains central in guiding and validating requirements.

\begin{figure}
    \centering
    \includegraphics[width=1.0\linewidth]{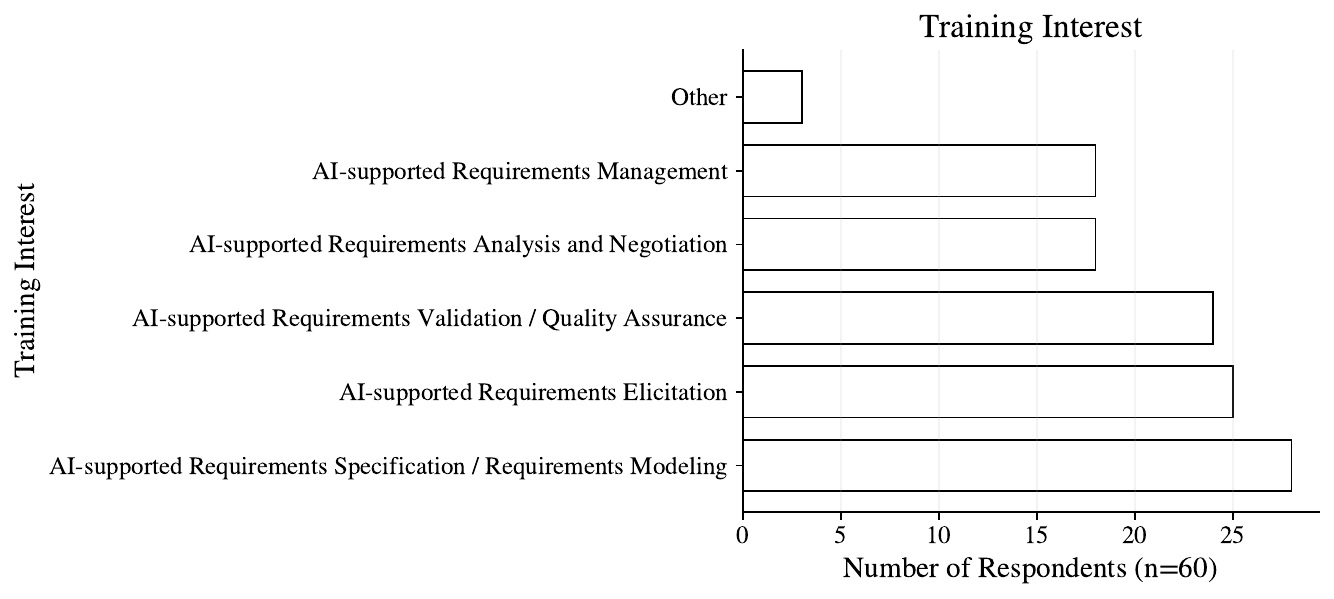}
    \caption{For which activities would you like to receive training?}
    \label{fig:training-interest}
\end{figure}

\begin{figure}
    \centering
    \includegraphics[width=1.0\linewidth]{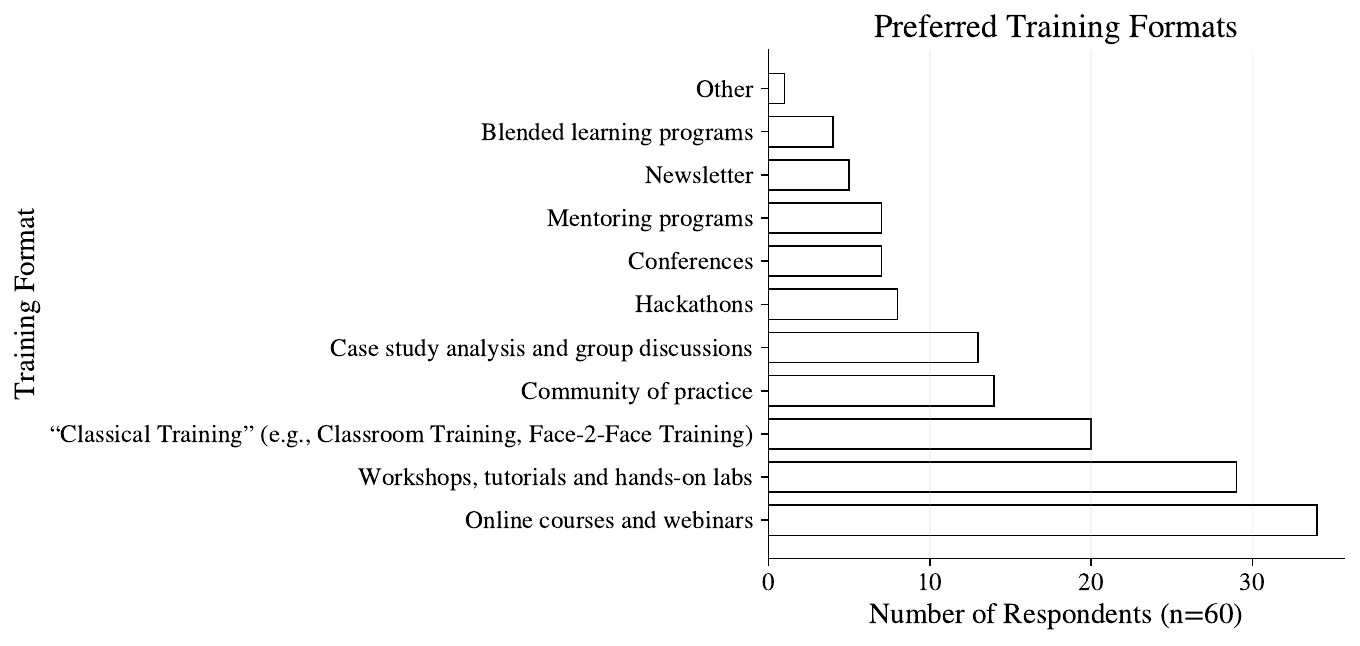}
    \caption{Which training format do you prefer? You can select up to three training formats.}
    \label{fig:training-format}
\end{figure}

Figure~\ref{fig:training-interest} shows that our respondents were interested in receiving training in all areas of RE. AI-supported requirements specification and modeling was selected most often.
Elicitation, validation, and QA were of similar interest. Negotiation and management were less frequently mentioned, but still selected by 30\% of the respondents. Respondents expressed a strong interest in training that helps them practically apply AI across all RE activities: 

\textbf{Elicitation}: using AI for prompting, analyzing documents and legacy data, and preparing interviews or questionnaires; in analysis and negotiation, supporting gap identification and structured discussions; 
\textbf{Specification and Modeling}: transforming unstructured inputs into well-structured requirements, diagrams, and models; 
\textbf{Validation and Quality Assurance}: using NLP, test generation, semantic checks, and conflict detection; 
\textbf{Management}: AI-supported impact analysis, prioritization, and tool integration. 

Across all phases, respondents also emphasized the importance of training on limitations, risks, and ethical considerations, as well as the practical integration of these concepts into existing processes and domain-specific contexts.
Figure~\ref{fig:training-format} shows that our respondents prefer online courses and webinars in addition to workshops, tutorials, and hands-on labs. Blended learning, newsletters, mentoring programs, conferences, and hackathons were less popular among the participants.

\section{Discussion}
\label{sec:Discussion}
\subsection{Comparison with Related Work}


According to our study, the RE activities most supported by AI in practice are primarily specification and modeling, as well as elicitation, followed by analysis and negotiation, validation and, lastly, requirements management. However, the literature reviews analyzed during our research indicate different emphases in academic publications. Table~\ref{tab:usageLit} compares the relative importance of these supported RE activities. In our survey, participants could select multiple RE activities, resulting in a cumulative response rate of 220 percent for ``yes'' answers. These responses are normalized to sum up to 100 in the table.

\newcolumntype{R}{>{\raggedleft\arraybackslash}X}
\begin{table}[t]
\centering
\caption{GenAI usage per RE activity in percent: comparison with related work}
\label{tab:usageLit}
\begin{tabularx}{\textwidth}{@{}lRRRR@{}}
\toprule
\textbf{RE activity}   & \textbf{Our study} & \textbf{Vasudevan and Reddivari}~\cite{ref_VR25}            & \textbf{Zadenoori et al.}~\cite{ref_ZDA25}*     & \textbf{Cheng et al.}~\cite{ref_CHL25}\\
\midrule
Elicitation   & 26.8 & 12.5 & 22 & 26.8 \\
Analysis and negotiation & 19.1 & 47.5 & 19 & 28.0 \\
Specification & 30.0 & 27.5 & 12 & 22.0 \\
Validation    & 17.7 & 10.0 & 22 & 17.9 \\
Management    & 6.4  & 2.5  & 12 & 5.4  \\
\bottomrule
\multicolumn{5}{l}{*not counting category ``other''}
\end{tabularx}
\vspace{-2em}
\end{table}

The six meta-studies~\cite{ref_CHL25,ref_CC25,ref_KDS24,ref_NCP25,ref_VR25,ref_ZDA25} analyzed conclude that the application of AI can accelerate RE activities and improve the quality of the results, while also bringing some challenges. Especially, humans are still needed to validate the semantical correctness of the results. This is consistent with our findings. However, these studies summarize mainly primary studies from science.


One study~\cite{ref_NCP25} considers all activities in software engineering, while another~\cite{molleri2020empirically} even all activities in engineering, including RE. These studies investigate research trends, highlighting which RE phases utilize LLMs/GenAI, alongside the models used, challenges faced, and future prospects. There have been two comparable surveys examining the adoption of generative AI in practical software engineering. However, these surveys also focused on GenAI usage at the broader level of software engineering phases, with RE being only one component~\cite{HIT2024,ref_KDS24}. In contrast, our survey not only explores which RE activities incorporate AI but also delves into specific RE usage scenarios within those activities, providing a more detailed perspective.

In our survey, we investigate the factors that hinder GenAI usage and the threats associated with its usage. Through the analysis of the six meta-studies~\cite{ref_CHL25,ref_CC25,ref_KDS24,ref_NCP25,ref_VR25,ref_ZDA25}, we found that these studies identified a similar mix of challenges, encompassing both preventing factors and threats, as our findings did. Key issues include concerns about confidential information data privacy~\cite{ref_CHL25,ref_CC25,ref_KDS24} and a lack of interpretability that affects stakeholder trust and limits control over the results generated~\cite{ref_CHL25,ref_CC25,ref_KDS24,ref_VR25}. Further challenges include the need for extensive high-quality datasets for AI training~\cite{ref_CC25,ref_KDS24,ref_VR25}, ensuring the accuracy and reliability of the output~\cite{ref_CC25,ref_KDS24,ref_NCP25}, and the integration of domain knowledge to refine AI~\cite{ref_CC25,ref_KDS24,ref_NCP25}. 
Limitations also involve the absence of advanced logical reasoning essential for complex problem-solving~\cite{ref_KDS24,ref_NCP25}, technical challenges such as computational power needs, and ethical concerns, including bias, fairness, and privacy~\cite{ref_CC25,ref_CHL25}. Additional issues relate to legal and regulatory compliance~\cite{ref_CC25,ref_CHL25}, integrating text and graphical data in various formats~\cite{ref_KDS24}, computation costs~\cite{ref_CHL25}, reproducibility of results \cite{ref_CHL25}, hallucinations \cite{ref_CHL25}, and the ambiguity and complexity of natural language~\cite{ref_CC25}. Other challenges include integrating AI into existing toolchains~\cite{ref_CC25}, the risk of over-reliance on AI leading to de-skilling of experts~\cite{ref_CC25}, ensuring security and robustness against malicious manipulations~\cite{ref_CC25}, the need for new user skills in AI utilization and interpretation \cite{ref_KDS24}, and the ongoing requirement to update and retrain models with the latest data~\cite{ref_KDS24}.

\subsection{Threats to Validity}
\textbf{Construct Validity.} There is a risk that respondents may have interpreted certain terms or questions differently. To mitigate this threat, we used commonly used terminology, explained key concepts, and conducted a pilot study in which participants reviewed the questionnaire for clarity and understandability.
To further reduce evaluation apprehension, the survey was conducted anonymously, without requiring any personal data.

\noindent\textbf{Content validity.} To mitigate the risk that the questions are not representative of what they aim to measure, participants in the pilot study were asked whether changes were needed to better capture the research questions. Nevertheless, we acknowledge that including additional questions could have allowed for a more comprehensive exploration of the phenomenon under investigation.

\noindent\textbf{Internal validity.} A potential threat to the study is the trustworthiness of responses. However, we did not find evidence to suggest that participants intentionally provided false answers. Another potential threat concerns the ability of the respondents' to accurately complete the questionnaire. We assumed that participants would have no difficulty using the LimeSurvey platform, which employs standard question formats, and that they would have sufficient English proficiency to understand the questions. To ensure clarity, we conducted a pilot study to verify the ease of comprehension of the questionnaire.

Since the study relies on self-reported data, responses can be influenced by self-serving and social desirability biases. Participants might overstate their adherence to recommended or modern practices or understate challenges. To mitigate these threats, the survey was anonymous and focused on concrete behaviors rather than evaluative judgments. The results should be interpreted as practitioners' perceptions rather than objective measurements.

This study is exploratory and descriptive. Given the non-random sampling and ordinal nature of the variables, the analysis was limited to descriptive statistics to avoid overinterpretation. More advanced statistical and demographic analysis are deferred to future studies.


\noindent\textbf{External validity.} The main threat pertains to the number and the representativeness of respondents and the projects or products they had experience with. Based on our study design and analysis of demographic data, the respondents represented a diverse set of software project participants (holding various roles, working in different organizations and industry sectors, etc.). This diversity helps mitigate the risk of bias toward a particular context. Although the sample appears appropriate for the objectives of the survey, we acknowledge that it may not be representative of all possible contexts, especially considering geographical diversity.
Sampling and self-selection bias were mitigated by distributing the survey through various professional channels, though voluntary participation may still limit generalizability.

\noindent\textbf{Conclusion Validity.} We tried to ensure that our qualitative analysis accurately reflected the data. Two researchers independently conducted qualitative coding, and two additional authors subsequently verified the correctness of the assigned codes. Discrepancies were discussed until consensus was reached, after which the final conclusions were drawn collaboratively.

\vspace{-1em}
\section{Conclusions}
\vspace{-1em}
\label{sec:Conclutions}
This paper presents the design and findings of an online survey with RE practitioners on their use and perception of GenAI. Most respondents primarily use GenAI for specification, modeling, and elicitation, with less frequent use in requirements management. Key barriers include ethical concerns, lack of awareness, insufficient support, and low-quality output. Time savings are a major benefit, but issues like hallucinations and over-reliance on AI necessitate careful review to ensure quality. Most respondents agreed that requirements engineers' skills must evolve as AI becomes more integrated into RE.

Our study contributes to the literature by focusing on GenAI use in RE activities and related usage scenarios from practitioners' perspectives. Positive experiences may inspire other practitioners to adopt GenAI in their work context. Researchers could replicate
this study to track the evolution of AI usage or to address current limitations by exploring regional differences with larger sample sizes. Finally, the results of the survey are intended to motivate the development of methods, tools and training programs to improve AI-supported RE activities by advancing skills in AI literacy, prompt engineering, and critical evaluation.

\begin{credits}
\vspace{1em}
\noindent\textbf{Data Availability Statement.}
The survey instrument (questionnaire), codebook, anonymized raw responses, and analysis materials supporting the findings of this study are publicly available via Zenodo at \url{https://doi.org/10.5281/zenodo.17429345}. All shared data were de-identified prior to publication to protect participant privacy.

\noindent\textbf{\ackname} We thank the respondents for sharing their valuable experience and the members of the SIG \#AIREB for their participation in the pilot study. This work was partially supported by Grant PID2024-156019OB-I00 funded by MICIU/AEI/10.13039/501100011033 and by ERDF, EU.

\noindent\textbf{Declaration of interest.} The authors have no competing interests to declare.
\vspace{-1em}

\end{credits}
%
%
%
\bibliographystyle{splncs04}
\bibliography{references}
\end{document}